\documentclass[a4paper,11pt]{article}
\pdfoutput=1
\usepackage{jheppub}

\usepackage{amssymb,amsmath}
\usepackage[normalem]{ulem}
\usepackage[utf8x]{inputenc}
\usepackage{slashed}
\usepackage{graphicx}
\usepackage{here}
\usepackage{color}
\usepackage{csquotes} 
\usepackage{comment}
\usepackage{mathrsfs}
\usepackage{float}
\usepackage{ascmac}
\usepackage{multirow}
\usepackage{longtable}
\usepackage{bm}
\usepackage{tcolorbox}

\usepackage[italicdiff]{physics}

\usepackage[hang,small,bf]{caption}
\usepackage[subrefformat=parens]{subcaption}
\newcommand{\Mpl}{M_\text{Pl}}

\makeatletter
\newcommand*\rel@kern[1]{\kern#1\dimexpr\macc@kerna}
\newcommand*\widebar[1]{%
  \begingroup
  \def\mathaccent##1##2{%
    \rel@kern{0.8}%
    \overline{\rel@kern{-0.8}\macc@nucleus\rel@kern{0.2}}%
    \rel@kern{-0.2}%
  }%
  \macc@depth\@ne
  \let\math@bgroup\@empty \let\math@egroup\macc@set@skewchar
  \mathsurround\z@ \frozen@everymath{\mathgroup\macc@group\relax}%
  \macc@set@skewchar\relax
  \let\mathaccentV\macc@nested@a
  \macc@nested@a\relax111{#1}%
  \endgroup
}
\makeatother

\numberwithin{equation}{section}

\preprint{
\begin{minipage}{5cm}
\small
\flushright
KYUSHU-HET-317
\end{minipage}}

\title{Higgs-Modular Inflation}

\author{Shuntaro Aoki$^{1}$,} 
\author{Hajime Otsuka$^{2}$, and} 
\author{Ryota Yanagita$^{2}$} 
\affiliation{
$^1$RIKEN Center for Interdisciplinary Theortical and Mathematical
Sciences (iTHEMS),
 Wako, Saitama 351-0198, Japan}
\affiliation{
$^2$Department of Physics, Kyushu University, 744 Motooka, Nishi-ku, Fukuoka 819-0395, Japan}
\emailAdd{shuntaro.aoki@riken.jp}
\emailAdd{otsuka.hajime@phys.kyushu-u.ac.jp}
\emailAdd{ryota.692@s.kyushu-u.ac.jp}

\abstract{
We investigate the role of the Higgs field as a fundamental scalar in the Standard Model within the framework of modular inflation models, where a modulus field acts as the inflaton and its interactions are governed by an underlying modular symmetry. In general, the Higgs field can participate in the dynamics of modular inflation, leading to a two-field inflationary system—termed \emph{Higgs-Modular inflation}—which exhibits non-trivial dynamics and interesting phenomenological implications. We analyze Higgs-Modular inflation both analytically and numerically, highlighting its attractor behavior and the resulting observational constraints. In particular, we find that Higgs-Modular inflation is favored by the latest data release from the Atacama Cosmology Telescope (ACT) in certain regions of parameter space. This is in contrast to both pure Higgs inflation and pure modular inflation with a Starobinsky-type potential, which tend to predict a relatively low spectral index. Additionally, we discuss the cutoff scale of this inflationary model and the reheating processes induced by the decays of the modulus and the Higgs field.}

\makeatletter
\gdef\@fpheader{}
\makeatother

\begin{document}

\maketitle

\section{Introduction}
Modular symmetry is well motivated in the context of string theory and has significant implications in both particle physics and cosmology. For example, it plays an essential role in flavor physics by helping to explain a unique pattern of flavor structure~\cite{Feruglio:2017spp}\footnote{See for reviews, Refs. \cite{Kobayashi:2023zzc,Ding:2023htn}.}, and in inflationary cosmology, where it supports the flatness of the potential and constrains interactions due to symmetry principles~\cite{Kobayashi:2016mzg,Schimmrigk:2016bde,Abe:2023ylh,Ding:2024neh,King:2024ssx}. Because of these symmetry-induced constraints, such frameworks are often predictive and provide a fertile ground for phenomenological studies.

In the study of flavor physics, it is commonly assumed that Standard Model (SM) fermions carry charges under the modular symmetry. Given that the Higgs field is a fundamental scalar in the SM, it is natural to extend this assumption and consider the possibility that the Higgs field also transforms under the modular symmetry. This raises important questions about its implications for both particle physics and cosmological phenomenology. In this work, we focus on the consequences of such a framework in the context of inflation, specifically investigating how the presence of the Higgs field, which carries the charge under the modular symmetry, affects inflationary dynamics. As a first step, we explore the impact of the Higgs field in a modular inflation scenario.

If the Higgs field carries a nonzero charge under the modular symmetry, it naturally couples to the modulus field that governs modular transformations. As a consequence, the Higgs field inevitably participates in inflationary dynamics, leading to a novel two-field inflationary scenario that we refer to as \textit{Higgs-Modular inflation}. The resulting dynamics involve nontrivial interactions between the Higgs and the modulus fields, significantly affecting the inflationary trajectory and observational predictions. We develop an analytic treatment of this system, whose validity we confirm by solving the full equations of motion numerically. Based on this analysis, we derive constraints on the allowed parameter space and discuss the observational signatures of the model. In particular, we find that Higgs-Modular inflation predicts a value of the spectral index \( n_s \) that falls within the favored region recently reported by the Atacama Cosmology Telescope (ACT)~\cite{ACT:2025fju, ACT:2025tim}, for certain choices of parameter space. In contrast, both pure Higgs inflation and pure modular inflation with a Starobinsky-type scalar potential tend to predict slightly lower values of \( n_s \).

So far, various extensions of Higgs inflation have been proposed and discussed in the literature, often motivated by UV completion (See Refs.~\cite{Giudice:2010ka,Bauer:2010jg,Barbon:2015fla,Ema:2017rqn,Shaposhnikov:2020gts,Aoki:2021aph,Aoki:2022csb} for example). In addition to the difference in inflationary observables such as the spectral index (and the tensor-to-scalar ratio) mentioned above, in order to further disentangle different models, it is important to study post-inflationary dynamics such as reheating. We examine the reheating dynamics within our Higgs-Modular inflation framework. We analyze how the presence of the Higgs field having the charge under the modular symmetry modifies the reheating process compared to conventional modular inflation scenarios, providing an additional handle for distinguishing different models.

The structure of this paper is as follows: In Sec.~\ref{Sec2}, we introduce the Higgs-Modulus interactions that are invariant under the $SL(2,\mathbb{Z})$ modular symmetry and present our setup for inflation. Then, in Sec.~\ref{Sec3}, we discuss the inflationary dynamics of Higgs-Modular inflation in detail, including predictions for observable quantities. We also study the cutoff scale in relation to unitarity issues. In Sec.~\ref{Sec4}, we analyze the post-inflationary dynamics, particularly the reheating process, by evaluating the decay rates of the modulus and Higgs fields and estimating the reheating temperature. Finally, Sec.~\ref{sec:con} is devoted to our summary and conclusions.

\section{Modular invariant Higgs-Modulus action}\label{Sec2}
Here we introduce modular invariant action particularly focusing on $SL(2,\mathbb{Z})$-invariant Higgs-Modulus couplings. The basics of non-holomorphic modular symmetry is summarized in Ref.~\cite{Qu:2024rns}.

\subsection{Modulus sector}\label{MS}
Under $SL(2,\mathbb{Z})$, the modulus $\tau$ transform as
\begin{align}
    \tau &\rightarrow \gamma \tau = \frac{p \tau + q}{r \tau +s}, \qquad 
    \gamma=
    \left(\begin{array}{ll}
p & q \\
r & s
\end{array}\right)
    \in SL(2,\mathbb{Z}),
\end{align}
and hence it follows
\begin{align}
    {\rm{Im} \tau}&\rightarrow \frac{ {\rm{Im} \tau}}{|r\tau +s|^2},
\quad
    \partial_\mu \tau \rightarrow \frac{\partial_\mu \tau}{(r\tau +s)^2}.
\end{align}
Then, the modular invariant kinetic term of $\tau$ is given by
\begin{align}
    \frac{\Mpl^2\partial^\mu \tau \partial_\mu \bar{\tau}}{(2a\rm{Im} \tau)^2} ,\label{tau_kin}
\end{align}
where the bar denotes the complex conjugate and we inserted a real parameter $a$ for generality.

Regarding the modulus self-interactions, the invariant modulus potential~\(V(\tau,\bar{\tau})\) has been discussed in the context of inflation in several works~\cite{Casas:2024jbw, Kallosh:2024ymt, Kallosh:2024pat,Aoki:2024ixq,Ding:2024euc}. As an example given in~\cite{Casas:2024jbw}, one may consider  
\begin{align}  
V=V_0\frac{|{\rm{Im}}\tau \tilde{G}_2|^2}{\left[\ln ({\rm{Im}}\tau |\eta(\tau)|^4)+N_0\right]^2},  
\end{align}  
where \( V_0 \) and \( N_0 \) are constants, and \( \tilde{G}_2 \) is an almost holomorphic modular form of weight \( 2 \), defined as  
\begin{align}  
\tilde{G}_2 \equiv -4\pi i \, \partial_\tau \ln \eta(\tau) - \frac{\pi}{{\rm{Im}}\tau},  
\end{align}  
with \( \eta(\tau) \) being the Dedekind eta function.

Another example is given in Refs.~\cite{Kallosh:2024ymt, Kallosh:2024pat}:  
\begin{align}  
V=V_0\left[1-\frac{\ln j(i)^2}{\ln \left(|j(\tau)|^2+j(i)^2\right)}\right],  \label{j_type}
\end{align}  
where \( j(\tau) = 12^3 J(\tau) \), with \( J(\tau) \) denoting Felix Klein's absolute invariant, such that \( J(i) = 1 \). Yet another example is given by  
\begin{align}  
V=V_0\left[1-\frac{\ln \eta^4(i)}{\ln ({\rm{Im}}\tau |\eta(\tau)|^4)}\right],  
\end{align}  
where \( \eta(\tau) \) is the Dedekind eta function.

The common feature of the above scalar potentials is, for ${\rm{Im}}\tau \gg 1$, they can be approximated as 
\begin{align}
V(\tau, \bar{\tau})\simeq  V_0\left(1-\frac{c}{{\rm{Im}}\tau}\right)^2,\label{asymp_V}
\end{align}
with $c\sim \mathcal{O}(1)$ being a model dependent number, and serve a flat potential suitable for inflation. 
The detailed analysis on the inflationary prediction of these models can be found in Refs.~\cite{Casas:2024jbw,Kallosh:2024ymt,Kallosh:2024pat, Aoki:2024ixq,Kallosh:2024whb,Carrasco:2025rud}.

Let us briefly comment on the origin of the modular potential discussed above. Such a modulus potential can arise in string compactifications. 
For instance, an \( SL(2, \mathbb{Z}) \)-invariant function appears in threshold corrections to gauge coupling constants~\cite{Dixon:1990pc}, which can induce a modulus potential when gauginos in the hidden gauge sector condense. 
See Ref.~\cite{Cvetic:1991qm} for a more general form of an \( SL(2, \mathbb{Z}) \)-invariant function.

\subsection{Higgs-Modulus interactions}
Next let us discuss Higgs-Modulus interactions,  which play a crucial role in our work. We assume that Higgs doublet $\mathbb{H}$ transforms under the finite modular group $\Gamma^\prime_N\equiv SL(2,\mathbb{Z})/\Gamma(N)$ or $\Gamma_N\equiv PSL(2,\mathbb{Z})/\Gamma(N)$ with $\Gamma(N)$ being a principal congruence subgroup of $SL(2,\mathbb{Z})$ as
\begin{align}
 &\mathbb{H} \rightarrow (r\tau +s)^{-k_{H}}\rho_{H} (\gamma) \mathbb{H}, \quad \mathbb{H}^\dagger  \rightarrow (r\bar{\tau} +s)^{-k_{H}}\rho^\dagger_{H} (\gamma) \mathbb{H}^\dagger,     
\end{align}
where $\rho_H$ and $k_H$ respectively denote the representation of the finite modular group and modular weight. Obviously, the Higgs kinetic term $\partial^{\mu}\mathbb{H}^{\dagger} \partial_{\mu}\mathbb{H}$ is not covariant under the above transformation. Therefore, we introduce the modular covariant derivative:
\begin{align}
    D_\mu = \partial_\mu + \frac{i \pi k_H }{6}E_2(\tau) \partial_\mu \tau,\label{cov}
\end{align}
where $E_2(\tau)$ is the weight 2 Eisenstein series:
\begin{align}
    E_2(\tau) = 1 - 24 \sum_{n=1}^\infty \sigma_1(n) q^n, 
\end{align}
with $q=e^{2\pi i \tau}$ and $\sigma_1(n) = \sum_{d|n} d$ being the sum of divisors of $n$. Since the Eisenstein series transforms as\footnote{The Eisenstein series itself is not a modular invariant form, but the following combination
\begin{align}
    \hat{E}_2(\tau) \equiv E_2(\tau) - \frac{3}{\pi {\rm Im}\tau},
\end{align}
is a polyharmonic Maa$\beta$ form of $SL(2,\mathbb{Z})$, i.e., $\hat{E}_2(\gamma \tau) = (r\tau +s)^2 \hat{E}_2(\tau)$.} 
\begin{align}
    E_2(\tau) \xrightarrow{SL(2,\mathbb{Z})} E_2(\gamma \tau) = (r\tau +s)^2 E_2(\tau) - \frac{6i}{\pi}r(r\tau +s),
\end{align} 
one can check that $D_\mu \mathbb{H}$ transforms in a covariant way:
\begin{align}
D_{\mu}\mathbb{H} \rightarrow (r\tau +s)^{-k_{H}}\rho_{H} (\gamma) D_{\mu}\mathbb{H}.    
\end{align}
Therefore, from the considerations above, we find that the following Higgs-Modulus couplings:  
\begin{align}  
\frac{ D^\mu \mathbb{H}^\dagger D_\mu \mathbb{H} }{(2{\rm{Im}}\tau)^{k_H}}, \quad    \frac{\mathbb{H}^\dagger \mathbb{H}}{(2{\rm{Im}}\tau)^{k_H}} \label{M-H-C}
\end{align}  
are invariant under \( SL(2,\mathbb{Z}) \), and thus can be included in the construction of the action.

\subsection{Total action}
Now, it is straightforward to write down Higgs-Modulus system in a modular invariant manner. In general, any function of Eq.~\eqref{M-H-C} can be modular invariant.
In this paper, we discuss the following simple Higgs-Modulus Lagrangian, 
\begin{align}
\mathcal{L}/\sqrt{-g}=\left[\frac{\Mpl^2}{2}+\xi \frac{\mathbb{H}^\dagger \mathbb{H}}{(2{\rm{Im}}\tau)^{k_H}}\right]R 
   -\frac{ D^\mu \mathbb{H}^\dagger D_\mu \mathbb{H} }{(2{\rm{Im}}\tau)^{k_H}}
    - \lambda\frac{(\mathbb{H}^\dagger \mathbb{H})^2}{(2{\rm{Im}}\tau)^{2k_H}}-\frac{\Mpl^2\partial^\mu \tau \partial_\mu \bar{\tau}}{(2\rm{Im} \tau)^2}-V(\tau,\bar{\tau}), \label{H_tau}
\end{align}
where $\xi$ and $\lambda$ are real coupling constants for the non-minimal coupling to gravity and the Higgs quartic term, respectively. One can also introduce the ``Higgs mass term'', $m^2\mathbb{H}^\dagger \mathbb{H}/(2{\rm{Im}}\tau)^{k_H} $, but it is irrelevant to the following discussion.

In summary, by allowing the Higgs to transform under the modular symmetry (i.e., assigning it a nonzero charge under the modular symmetry), we obtain a Higgs-modulus system as described in Eq.~\eqref{H_tau}. It is interesting to derive the above Higgs-Modulus system from a certain UV theory\footnote{See, e.g., Ref.~\cite{Kikuchi:2022txy}, for a realization of modular symmetric models from higher dimensional theory.}, such as string theory, but it is beyond the scope of this paper. Our primary focus is to investigate how the presence of the Higgs field influences the modular inflation scenario, which we will explore in detail in the next section.

\section{Inflationary analysis}\label{Sec3}
In this section, we discuss inflationary dynamics of Higgs-Modulus system given in Eq.~\eqref{H_tau}. In the following, we set $\Mpl=1$ unless explicitly stated.

\subsection{Einstein frame Lagrangian}
Our action~\eqref{H_tau} is given in the Jordan frame, where the Ricci scalar is not canonically normalized. To discuss the inflationary dynamics, it is convenient to move to the Einstein frame. To begin with, let us rewrite Eq.~\eqref{H_tau} in the unitary gauge, \( \mathbb{H} = (0, h/\sqrt{2}) \), 
\begin{align}
\nonumber \mathcal{L}_J/\sqrt{-g_J}=&\ \frac{1+\Xi(\tau,\bar{\tau}) h^2}{2} R_J -\frac{\mathcal{H}(\tau,\bar{\tau})}{2}g_J^{\mu\nu} D_\mu h \left(D_\nu h\right)^{\dagger}-\frac{\mathcal{K}(\tau,\bar{\tau})}{2}g_J^{\mu\nu} \partial_\mu \tau \partial_\nu \bar{\tau}\\
&-\left[\frac{\Lambda(\tau,\bar{\tau})}{4}h^4+V_J(\tau,\bar{\tau})\right],
\end{align}
where the subscript ``J'' on several quantities explicitly indicates that the action is given in Jordan frame. 
We also introduced modulus dependent functions,
\begin{align}
\Xi(\tau,\bar{\tau}) \equiv \frac{\xi}{(2{\rm Im}\tau )^{k_H}},\quad    \mathcal{H}(\tau,\bar{\tau}) \equiv \frac{1}{(2{\rm Im}\tau )^{k_H}}, \quad    \mathcal{K}(\tau,\bar{\tau}) \equiv \frac{1}{(2a{\rm Im}\tau )^{2}},\quad \Lambda(\tau,\bar{\tau}) \equiv \frac{\lambda}{(2{\rm Im}\tau )^{2k_H}}, \label{func_tau}
\end{align}
where note that $\Lambda$ is irrelevant to the cutoff scale of the theory; do not confuse this.

Moving to Einstein frame can be achieved by 
the following conformal transformation
\begin{align}
g_{J, \mu\nu}=\Omega^{-2}g_{E, \mu\nu}, \quad   \Omega^2 \equiv 1+\Xi(\tau,\bar{\tau})h^2,
\end{align}
with some useful formulae
\begin{align}
&\sqrt{-g_J}=\Omega^{-4} \sqrt{-g_E},\label{formulae1}\\
&R_J=\Omega^2 R_E-\frac{3}{2} \Omega^2g_E^{\mu\nu}\partial_\mu \log \Omega^2\partial_\nu \log \Omega^2+3 \Omega^2 g_E^{\mu\nu}\partial_\mu\partial_\nu \log \Omega^2,\label{formulae2}
\end{align}
and we obtain 
\begin{align}
\mathcal{L}_E/\sqrt{-g_E}=&\ \frac{R_E}{2}  -\frac{\mathcal{H}(\tau,\bar{\tau})}{2\Omega^2}|D_\mu h|^2-\frac{\mathcal{K}(\tau,\bar{\tau})}{2\Omega^2}|\partial_\mu \tau|^2 -\frac{3}{4} (\partial_\mu \log \Omega^2)^2-V_E,  \label{E_action} 
\end{align}
where we use abbreviation $|D_\mu h|^2=g_E^{\mu\nu} D_\mu h \left(D_\nu h\right)^{\dagger}$, and the scalar potential in Einstein frame is given by 
\begin{align}
V_E\equiv \frac{1}{\Omega^4}\left[\frac{\Lambda(\tau,\bar{\tau})}{4}h^4+V_J(\tau,\bar{\tau})\right]. \label{P_E}   
\end{align}
In the following discussion, we will omit the subscript ``E'' on the metric. The action~\eqref{E_action} contains three 
 scalar degree of freedom: 
 \begin{align}
 &{\rm{Higgs}}: h,\\
 &{\rm{Moduls}}: \tau = {\rm{Re}}\tau+i  {\rm{Im}}\tau \equiv \tau_1+i\tau_2. 
 \end{align}
In this paper, we focus on a two-field system described by \(\{h, \tau_2\}\), setting \(\tau_1 = 0\), while leaving a detailed analysis of the dynamics of \(\tau_1\) for future study.\footnote{In general, the \(\tau_1\)-direction is very light in the type of scalar potentials discussed in Sec.~\ref{MS}, which may raise concerns about the enhancement of isocurvature perturbations. A detailed discussion on the isocurvature mode can be found in \cite{Kallosh:2024whb}. Additionally, Ref.~\cite{Carrasco:2025rud} explores possible mechanisms to make \(\tau_1\) heavy without altering the inflationary dynamics or violating \( SL(2,\mathbb{Z}) \)-symmetry. For instance, the \( j \)-type scalar potential~\eqref{j_type} can be modified as  
\begin{align}  
V=V_0\left[1-\frac{\ln j(i)^2}{\ln \left(|j(\tau)|^2+A|j(\tau)-\overline{j(\tau)}|^2+j(i)^2\right)}\right],  
\end{align}  
where the correction term proportional to \( A \) provides mass to \(\tau_1\). We naively expect that this mechanism also applies to our case, including the Higgs field \( h \), provided that \(\xi h^2\) is not too large.}

Taking into account that the Higgs covariant derivative~\eqref{cov} can be approximated as  
\begin{align}
|D_\mu h|^2\simeq (\partial_\mu h)^2 -\frac{k_H \pi}{3}h \partial_\mu h \partial^\mu \tau_2 +\frac{k_H^2\pi^2}{36}h^2(\partial_\mu \tau_2)^2,\label{cov_h}
\end{align}
for $\tau_2 \gtrsim 1$, our two-field system can be summarized as
\begin{align}
\mathcal{L} / \sqrt{-g}=&\frac{R}{2} -\frac{1}{2}G_{ab}\partial_\mu \phi^a\partial^\mu \phi^b -V_E,\label{double_system}
\end{align}
where $\phi^a=\{\tau_2, h\}$ and the field space metric $G_{ab}$ is given by
\begin{align}
&G_{\tau_2\tau_2}=\frac{\mathcal{K}}{\Omega^2}+\frac{3\Xi_{\tau_2}^2h^4}{2\Omega^4}+\frac{k_H^2\pi^2\mathcal{H}h^2}{36\Omega^2},\\
&G_{\tau_2h}=\frac{3\Xi_{\tau_2}\Xi h^3}{\Omega^4}-\frac{ k_H\pi\mathcal{H}h }{6\Omega^2},\\
&G_{hh}=\frac{\mathcal{H}}{\Omega^2}+\frac{6\Xi^2h^2}{\Omega^4},
\end{align}
with $\Xi_{\tau_2}\equiv \partial \Xi/\partial {\tau_2}$. We note that $\tau_2$ and $h$ interact even when the modular weight of the Higgs is zero ($k_H = 0$). However, a nonzero modular weight ($k_H \neq 0$) significantly modifies the interactions. The equations of motion can be derived from Eq.~\eqref{double_system}, 
\begin{align}
&D_t \dot{\phi}^a+3 H \dot{\phi}^a+G^{ab} V_b=0,\label{EOM1}\\
&3  H^2=\frac{1}{2} \dot{\phi}^2+V,\label{EOM2}\\
& \dot{H}=-\frac{1}{2} \dot{\phi}^2,\label{EOM3}
\end{align}
where the dot denotes the time derivative and we define $D_t \dot{\phi}^a \equiv \ddot{\phi}^a+\Gamma_{bc}^a \dot{\phi}^b \dot{\phi}^c$, $V_a\equiv \partial V/\partial \phi^a$ and $\dot{\phi}^2\equiv G_{ab}\dot{\phi}^a\dot{\phi}^b$.

Before going to study Eq.~\eqref{double_system} in detail, we can roughly classify the situations as follows:  
\begin{itemize}
    \item {\bf Pure Higgs inflation regime:} In the limit where the modulus potential dominates in Eq.~\eqref{P_E}, i.e.,  $V_J\gg \Lambda h^4/4$, the scalar potential is minimized at $\tau_2={\rm{const.}}$ and one recovers the original Higgs inflation~\cite{Bezrukov:2007ep}. The inflationary observables (such as spectral index $n_s$ and tensor-to-scalar ratio $r$) are predicted by 
    \begin{align}
    n_s \simeq  1 -\frac{2}{N}-\frac{9}{2 N^2}, \quad r \simeq \frac{12}{N^2},   \label{nsr_Higgs} 
    \end{align}
    which is consistent with Planck data~\cite{Planck:2018jri} but slightly disfavored by ACT results as pointed out in Ref.~\cite{Kallosh:2025rni}. 

    \item {\bf Pure modular inflation regime:} On the other hand, in the opposite limit with $V_J\ll \Lambda h^4/4$, the potential is minimized at $h=0$ and the pure modular inflation with $V_E\simeq V_J(\tau_2)$ is realized. The modular inflation scenario with several types of potential forms is discussed in Refs.~\cite{Casas:2024jbw,Kallosh:2024ymt,Kallosh:2024pat}, and some scalar potential forms $V_J(\tau_2)$ are briefly summarized in Sec.~\ref{MS}. For $\tau_2\gg 1$, their asymptotic forms are given by Eq.~\eqref{asymp_V}, which is similar to the Starobinsky-type potential, giving the same observational prediction as Eq.~\eqref{nsr_Higgs}.

 \item {\bf Mixed regime:} Finally, in the intermediate region where both contributions are compatible $V_J\sim \Lambda h^4/4$, it realizes Higgs-Modulus mixed inflation. This is the main topic of the paper, which we investigate in detail in the next subsection.

\end{itemize}

\subsection{Higgs-Modulus mixed inflation}
\label{sec:modular_inf}
Let us study the mixed regime $V_J\sim \Lambda h^4/4$ classified in the previous subsection, where both modulus and Higgs play important roles.

In general, the analysis of two-field inflation requires a numerical study of Eqs.~\eqref{EOM1}-\eqref{EOM3}, since the dynamics depends on the initial conditions and does not exhibit attractor behavior. In our case, however, there exists an approximate trajectory which can be obtained by  
\begin{align}
\frac{\partial V_E}{\partial h}=0, \quad \Rightarrow \quad h^2= \begin{cases}\frac{4\xi(2\tau_2)^{k_H}}{\lambda}V_J(\tau_2),  & \xi>0, \\0, & \xi<0.\end{cases} 
\label{eq:h_VEV}   
\end{align}
Note that the negative non-minimal coupling yield pure modular inflation scenario and therefore we do not consider this situation in the following, assuming $\xi >0$. Inserting Eq.~\eqref{eq:h_VEV} into the original action~\eqref{double_system}, we obtain a single field effective description along the trajectory 
\begin{align}
\mathcal{L}_{\rm{eff}}/\sqrt{-g}\simeq -\frac{1}{2}\frac{\lambda+4a^2k_H^2\xi V_J(1-\pi \tau_2/3)^2}{4a^2\left(\lambda +4\xi^2V_J\right)\tau_2^2}
(\partial_\mu \tau_2)^2-\frac{\lambda  V_J}{\lambda +4 \xi^2 V_J}, \label{SEFT}
 \end{align}
where we ignored terms suppressed by slow-roll parameters such as $\partial_{\tau_2} V_J$.

To study the inflationary predictions of Eq.~\eqref{SEFT}, we assume the form of the modulus potential as given in Eq.~\eqref{asymp_V}, which is repeated here:  
\begin{align}  
V_J\left(\tau_2\right) = V_0 \left(1 - \frac{c}{\tau_2} \right)^2,  
\end{align} \label{VJ} 
motivated by the asymptotic behavior of several types of modular inflation models that are consistent with observations (see examples in Sec.~\ref{MS}). Although the potential is specified, we emphasize that the effective action~\eqref{SEFT} is valid independently of the form of the moduli potential, as long as it exhibits a plateau, \( V_J(\tau_2) \approx V_0 \), during inflation.

Before proceeding, let us confirm the stability of the trajectory~\eqref{eq:h_VEV}. To do so, let us change the field basis from $\{\tau_2,h\}$ to $\{\tau_2,\chi\}$ by
\begin{align}
\chi \equiv \frac{\xi h^2}{(2\tau_2)^{k_H}}.    \label{eq:chi_VEV}
\end{align}
Note that the trajectory~\eqref{eq:h_VEV} is now given by $\chi=4\xi^2V_J/\lambda\simeq {\rm{const.}}$. In this field basis, the kinetic matrix reads
\begin{align}
&G_{\tau_2\tau_2}=\frac{
 9 \xi+a^2 k_H^2 (3 - \pi \tau_2)^2 \chi 
}{
36 a^2 \xi \tau_2^2 (1+\chi )
},\\
&G_{\tau_2\chi}=\frac{
 k_H(3 - \pi \tau_2)
}{
12 \xi \tau_2 (1+\chi) 
},\\
&G_{\chi\chi}=\frac{6 \xi  \chi +\chi +1}{4 \xi  \chi  (1+\chi )^2}.
\end{align}
Therefore, for $\chi\lesssim 1$, the off-diagonal element is suppressed and it is enough to focus on the mass matrix. Along the trajectory~\eqref{eq:chi_VEV}, the (canonical) mass of $\chi$ is evaluated as
\begin{align}
m^2_{\chi}\simeq \frac{8\lambda^2\xi V_0\Mpl^2}{\left(\lambda+4\xi^2 V_0\right)\left(\lambda+4\xi^2(1+6\xi) V_0\right)}\simeq \frac{24 \lambda \xi H^2}{\lambda+4\xi^2(1+6\xi) V_0}, 
\end{align}
where we used a relation $3\Mpl^2H^2\simeq \lambda V_0\Mpl^4/(\lambda+4\xi^2 V_0)$ during inflation. 
Therefore, for $\lambda \gtrsim\xi^3 V_0$ and $\xi\gtrsim 1$, the $\chi$-mass is sufficiently heavy compared to the Hubble scale and one can focus on the single field theory~\eqref{SEFT}.

Generally, we can divide Eq.~\eqref{SEFT} into two situations depending on the choice of parameters.

\subsubsection*{Case I (Modular-like)}
First, we focus on the regime where $\tau_2$ satisfies  
\begin{align}
    \lambda \gg k_H^2 \xi V_J \tau_2^2, \label{cond_1}
\end{align}
in addition to $\tau_2 \gg 1$, during inflation. 
Note that this condition is automatically satisfied when $k_H=0$ (Higgs is not charged under modular symmetry). In this case, the kinetic coefficient in Eq.~\eqref{SEFT} can be approximated by $\sim \lambda/4a^2(\lambda +4\xi^2V_J)\tau_2^2$, and then,
a canonically normalized inflaton $\phi$ is defined by 
\begin{align}
\phi= \frac{\log \tau_2}{\alpha}, \quad {\rm {with}}  \quad \alpha\equiv 2a \sqrt{\frac{\lambda+4\xi^2V_0}{\lambda}}.
\end{align}
The scalar potential for $\phi$ reads
\begin{align}
V(\phi)\simeq \frac{\lambda V_0\left(1-c e^{-\alpha \phi}\right)^2}{\lambda+4 \xi^2 V_0\left(1-c e^{-\alpha \phi}\right)^2}, \label{V_inf}
\end{align}
which takes the similar form as $\alpha$-attractor model~\cite{Kallosh:2013yoa}.

Given the potential~\eqref{V_inf}, it is straightforward to estimate inflationary observable. The potential slow-roll (SR) parameters are evaluated as 
\begin{align}
&\epsilon \equiv\frac{1}{2}\left(\frac{d V / d \phi}{V}\right)^2 \simeq \frac{8a^2 c^2 \lambda }{\lambda+4 \xi^2 V_0} e^{-2 \alpha \phi},\label{epsilon}\\
&\eta \equiv \frac{d^2 V / d \phi^2}{V}\simeq -8a^2 c e^{- \alpha \phi},\label{eta}
\end{align}
for $\phi \gtrsim 1$. The number of e-folding $N$ is calculated 
\begin{align}
N=\int_{\phi_e}^{\phi} \frac{1}{\sqrt{2 \epsilon}} d \phi \simeq e^{\alpha \phi}/8a^2c=\tau_2/8a^2c,  \label{N_phi}  
\end{align}
where $\phi_e$ is the field value at the end of inflation and we used $\phi\gg \phi_e$ in the second equality. By substituting the above relation between $\phi$ and $N$ into the SR parameters in Eqs.~\eqref{epsilon} and~\eqref{eta}, we obtain
\begin{align}
\epsilon \simeq \frac{1}{2\alpha^2N^2},\quad \eta\simeq -\frac{1}{N}.
\end{align}
Therefore, one can express inflationary observable (spectral index $n_s=1-6\epsilon+2\eta$ and tensor-to-scalar ratio $r=16\epsilon$) by the e-folding $N$ as
\begin{align}
n_s \simeq  1- \frac{2}{N}-\frac{3}{\alpha^2 N^2 }, \quad r \simeq \frac{8}{\alpha^2N^2}.    \label{ns_r_1}
\end{align}
The prediction is almost identical to that of the pure Higgs inflation model as well as the pure modulus inflation model (see Eq.~\eqref{nsr_Higgs}) for \( \alpha \sim \mathcal{O}(1) \), which holds unless \( a \) is extremely small.
Finally, the amplitude of the power spectrum $A_s$ is evaluated as
\begin{align}
A_s\equiv  \frac{V}{24 \pi^2\epsilon}\simeq \frac{a^2N^2V_0}{3\pi^2},\label{A_s}
\end{align}
at the leading order, which is fixed by the CMB normalization: 
\begin{align}
  A_s=  2.1 \times 10^{-9}, \quad \Rightarrow \quad V_0= \frac{1}{a^2}\left(\frac{60}{N}\right)^2 \times 1.7\times 10^{-11}.\label{V_0}
\end{align}
Note that the result~\eqref{A_s} is independent of Higgs quartic $\lambda$ and non-minimal coupling $\xi$.

\begin{figure}[t]
\centering
\begin{minipage}[b]{0.32\columnwidth}
    \centering
    \includegraphics[width=1\columnwidth]{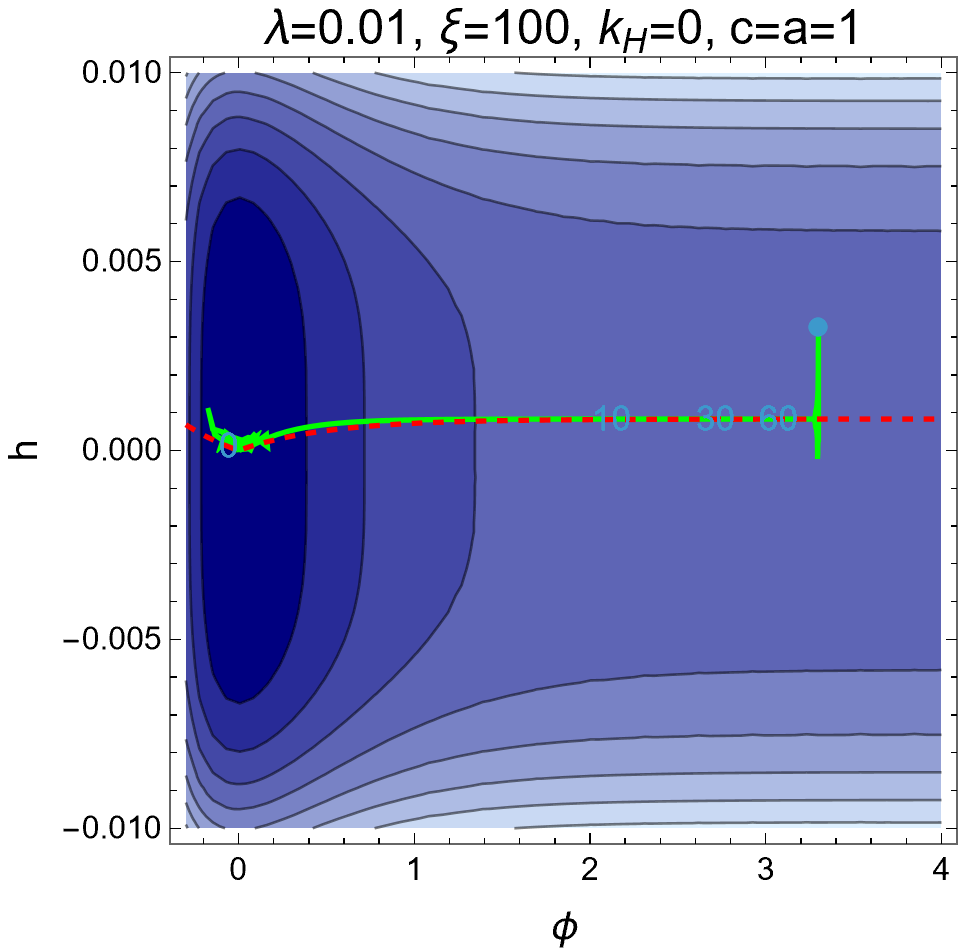}
\end{minipage}
\begin{minipage}[b]{0.32\columnwidth}
    \centering
    \includegraphics[width=1\columnwidth]{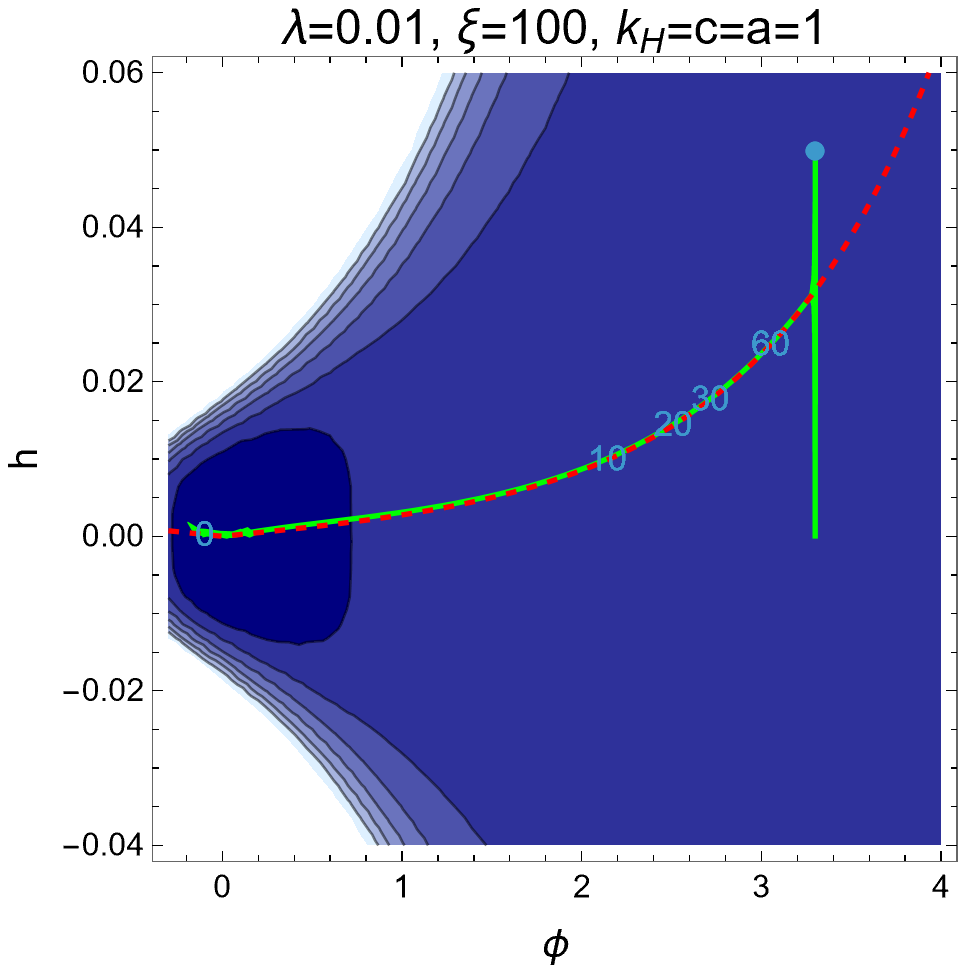}
\end{minipage}
\begin{minipage}[b]{0.32\columnwidth}
    \centering
    \includegraphics[width=1\columnwidth]{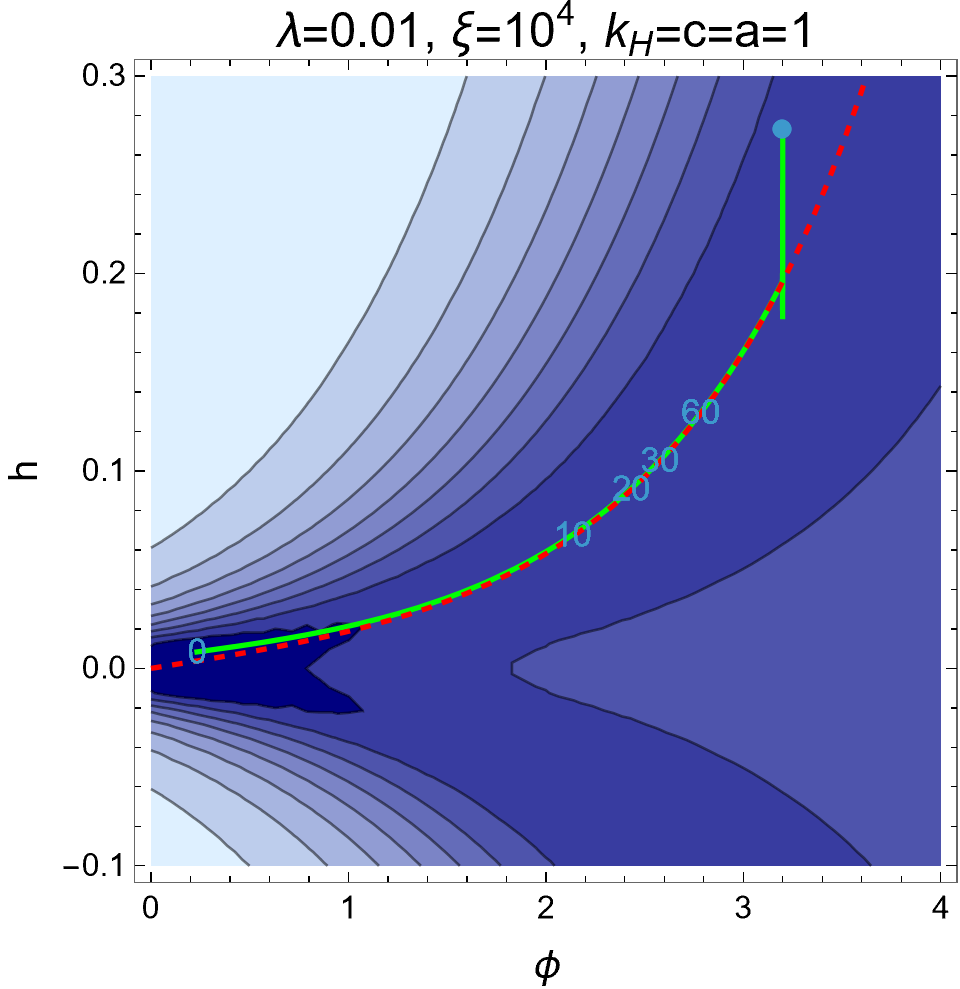}
\end{minipage}
\caption{The time evolution of the Higgs-Modular system (\(\phi\): modulus and \(h\): Higgs) is shown, where the smaller the value of the scalar potential, the darker the color is.
The green line represents the full numerical solutions, while the red dashed line corresponds to the approximated trajectory given by Eq.~\eqref{eq:h_VEV}. Several e-folds are marked along the trajectory. The parameters are chosen as indicated in the panel. The value of \( V_0 \) is fixed by the CMB normalization. }
  \label{fig:NS}
\end{figure}

In Fig.~\ref{fig:NS}, we show the results of the inflationary evolution (green line) for several parameter choices by numerically solving the full equations of motion~\eqref{EOM1}-\eqref{EOM3}. The two figures from the left correspond to the case I satisfying Eq.~\eqref{cond_1} at $N\sim 60$. The red dashed line corresponds to the analytically obtained trajectory~\eqref{eq:h_VEV}. One can see that this trajectory describes the dynamics well. Even though we set the initial conditions apart from the trajectory, it reaches the trajectory after a few e-foldings following oscillations in the Higgs direction and then settles down, indicating that the trajectory is an attractor.

Using Eq.~\eqref{N_phi} with $N=60$, the condition~\eqref{cond_1} can be estimated as 
\begin{align}
 \lambda/\xi \gg 1.7\times 10^{-5} ,\label{case_I}
\end{align}
where we set $a=c=k_H=1$ and $V_0$ is set by Eq.~\eqref{V_0}.
This yields the validity range of parameters that inflationary predictions~\eqref{ns_r_1} are reliable.

\subsubsection*{Case II (Higgs-like)}
In the opposite regime to Eq.~\eqref{cond_1}, i.e.,
\begin{align}
\lambda\ll k_H^2\xi V_J \tau_2^2,\label{cond_2}
\end{align}
during inflation, the kinetic coefficient of Eq.~\eqref{SEFT} is approximated as constant, and the canonically normalized field $\psi$ is given by
\begin{align}
\psi \equiv \frac{\pi k_H}{3}\sqrt{\frac{\xi V_0}{\lambda+4\xi^2V_0}}\ \tau_2.  
\end{align}
In the same way as the previous calculation, we can obtain the relation between the e-folding $N$ and the field value $\psi$ or $\tau_2$,
\begin{align}
N\simeq \frac{\psi^3}{2\pi ck_H\lambda}\sqrt{\frac{(\lambda+4\xi^2V_0)^3}{\xi V_0}}=\frac{\pi^2k_H^2\xi V_0 \tau_2^3}{54c\lambda},\label{N_psi}
\end{align}
for $\tau_2 \gg 1$, and based on this, the inflationary observable $(n_s,r)$ are estimated as 
\begin{align}
&n_s\simeq 1-\frac{4}{3N}- \frac{(4\pi^2 c^2 k_H^2 \lambda^2\xi V_0)^{1 / 3}}{3 N^{4 / 3}\left(\lambda+4 \xi^2 V_0 \right)},\label{n_s_psi}\\
&r\simeq \frac{8(4\pi^2 c^2 k_H^2 \lambda^2\xi V_0)^{1 / 3}}{9 N^{4 / 3}\left(\lambda+4 \xi^2 V_0 \right)}.\label{r_psi}
\end{align}
The CMB normalization fixes $V_0$ as
\begin{align}
V_0=\sqrt{\frac{\xi}{\lambda}}ck_H\left(\frac{60}{N}\right)^2\times 8.0\times 10^{-15}.    \label{V_02}
\end{align}
Note that the potential form in Case II, \( V_J(\psi) \sim (1-\mathcal{C}/\psi)^2 \), with a constant \( \mathcal{C} \), and hence its observational predictions~\eqref{n_s_psi} and \eqref{r_psi} resemble those of the brane inflation scenario~\cite{Burgess:2001fx}. This feature is specific to our Higgs–modular mixed inflation scenario, which differs from the polynomially flat modulus potentials considered in the literature~\cite{Ding:2024neh}.

In the same way as in Case I, evaluating Eq.~\eqref{cond_2} with Eq.~\eqref{N_psi} at the CMB scale gives the opposite inequality to Eq.~\eqref{case_I}, when the parameters are fixed by Eq.~\eqref{V_02}.  

\begin{figure}[t]
\centering
\begin{minipage}[b]{0.4\columnwidth}
    \centering
    \includegraphics[width=1\columnwidth]{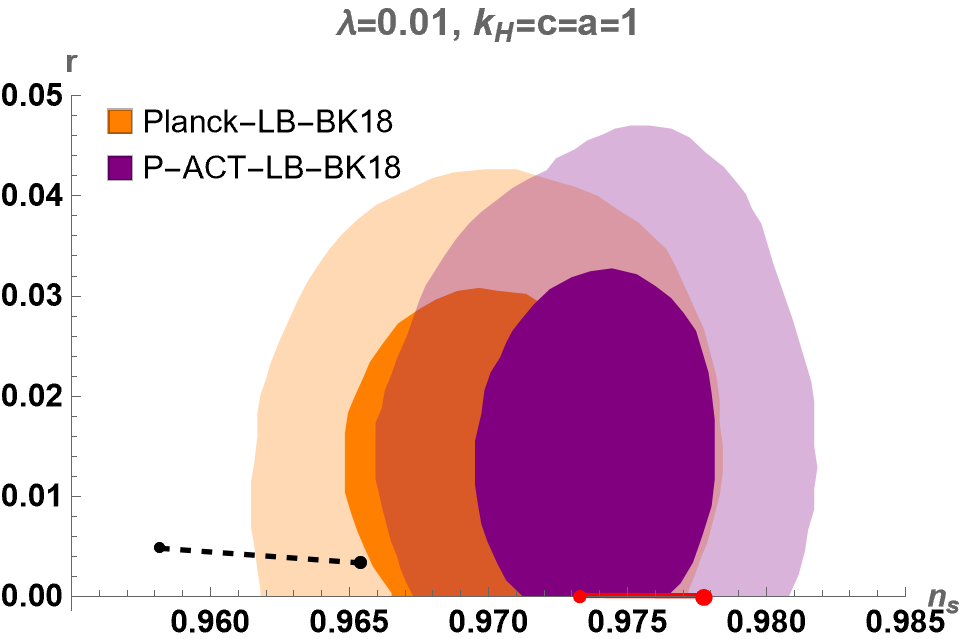}
\end{minipage}
\begin{minipage}[b]{0.4\columnwidth}
    \centering
    \includegraphics[width=1\columnwidth]{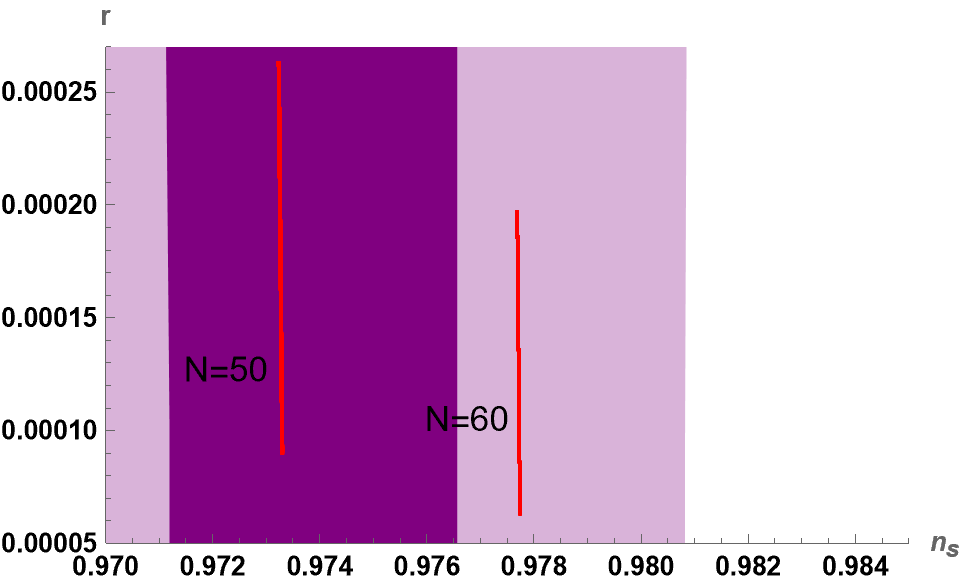}
\end{minipage}
\caption{{\it{Left:}} The constraints on \( n_s \) and \( r \), derived from the combined results of Planck, ACT, and DESI (P-ACT-LB-BK18) data~\cite{ACT:2025fju}, are shown as dark and light purple regions, corresponding to the 1\(\sigma\) and 2\(\sigma\) confidence levels, respectively. The constraints from Planck-LB-BK18 are shown in dark and light orange. The black dashed line represents the prediction of Higgs inflation, where the left (right) edge corresponds to the number of e-folds \(N = 50 (60)\). The red line represents the prediction of Higgs-Modular inflation, with parameters chosen as indicated. {\it{Right:}} A magnified view of the left figure. The non-minimal coupling \( \xi \) varies within the range \( \lambda/(1.7\times 10^{-5}) \leq \xi \leq 10^4 \), represented by the red line.
}
  \label{fig:YB} 
\end{figure}

The right panel of Fig.~\ref{fig:NS} corresponds to Case II. Similar to Case I, the trajectory given by Eq.~\eqref{eq:h_VEV} accurately describes the dynamics during inflation. Additionally, the left panel of Fig.~\ref{fig:YB} presents the \( n_s \)-\( r \) plane for Case II (red), along with the prediction of pure Higgs inflation (or pure Modular inflation with a Starobinsky-type potential~\eqref{asymp_V}) for comparison (black). We find that Case II predicts a relatively high spectral index, which is favored by the P-ACT-LB-BK18 data. The right panel of Fig.~\ref{fig:YB} shows a magnified view of the left one, where the non-minimal coupling \( \xi \) varies within the range \( \lambda/(1.7\times 10^{-5})\leq\xi \leq 10^4 \), represented by the red line. We present results for both \( N=50 \) and \( N=60 \), finding that the predictions for \( N=50 \) fall within the \( 1\sigma \) confidence region.

\subsection{Unitarity}
Here, let us briefly discuss the cutoff scale of Higgs-Modular inflation. In pure Higgs inflation, this has been extensively discussed in the context of unitarity issues~\cite{Burgess:2009ea, Barbon:2009ya, Burgess:2010zq, Hertzberg:2010dc}, so we are interested in examining what happens in our case.

First, without taking the unitary gauge, the Einstein frame action~\eqref{E_action} can be summarized as
\begin{align}
\mathcal{L} / \sqrt{-g}=\frac{R}{2}-\frac{G_{ab}}{2}g^{\mu\nu}\partial_\mu \varphi^a\partial_\nu \varphi^b -V_E,
\end{align}
where $\varphi^a=\{\tau_1,\tau_2,\phi_i\}$ with $ \tau_1\equiv {\rm{Re}}\tau $, $\tau_2 \equiv {\rm{Im}}\tau$, and  $\phi_i\ (i=1,2,3,4)$ being component of Higgs field. The field space metric $G_{ab}$ is given by
\begin{align}
    G_{ab}&=
\begin{pmatrix}
    A & B \\ B^{T} & C
\end{pmatrix},
\end{align}
with
\begin{align}
   & A =
    \begin{pmatrix}
        \frac{\mathcal{K}}{\Omega^2}+\frac{k_H^2\pi^2\mathcal{H}\phi^2}{36\Omega^2} & 0 \\ 0 & \frac{\mathcal{K}}{\Omega^2}+\frac{3}{2}\left(\frac{\Xi_{\tau_2}\phi^2}{\Omega^2}\right)^2+\frac{k_H^2\pi^2\mathcal{H}\phi^2}{36\Omega^2}
    \end{pmatrix},\\
   & B =
     \begin{pmatrix}
        -\frac{k_H\pi\mathcal{H}\phi_2}{6\Omega^2} & \frac{k_H\pi\mathcal{H}\phi_1}{6\Omega^2} & -\frac{k_H\pi\mathcal{H}\phi_4}{6\Omega^2} & \frac{k_H\pi\mathcal{H}\phi_3}{6\Omega^2} \\ \frac{3\Xi\Xi_{\tau_2}\phi^2\phi_1}{\Omega^4}-\frac{k_H\pi\mathcal{H}\phi_1}{6\Omega^2} & \frac{3\Xi\Xi_{\tau_2}\phi^2\phi_2}{\Omega^4}-\frac{k_H\pi\mathcal{H}\phi_2}{6\Omega^2} & \frac{3\Xi\Xi_{\tau_2}\phi^2\phi_3}{\Omega^4}-\frac{k_H\pi\mathcal{H}\phi_3}{6\Omega^2} & \frac{3\Xi\Xi_{\tau_2}\phi^2\phi_4}{\Omega^4}-\frac{k_H\pi\mathcal{H}\phi_4}{6\Omega^2}
    \end{pmatrix},\\
 &   C = \frac{\mathcal{H}}{\Omega^2}\delta_{ij}+\frac{6\Xi^2\phi_i\phi_j}{\Omega^4}.
\end{align}
Here we define $\Omega^2 \equiv 1+\Xi(\tau,\bar{\tau})\phi^2$ and $\phi^2=\sum_i\phi_i^2$. We remind moduli dependent functions $\mathcal{K},\mathcal{H},\Xi$ are given in Eq.~\eqref{func_tau}. Finally, the scalar potential $V_E$ is given by
\begin{align}
 V_E\equiv \frac{1}{\Omega^4}\left[\frac{\Lambda(\tau,\bar{\tau})}{4}(\phi^2)^2+V_J(\tau,\bar{\tau})\right].   
\end{align}

To estimate the cutoff scale of the Higgs-Modular system, we compute the Ricci scalar constructed from \( G_{ab} \), ignoring potential effects. We note that this is a simplified argument since the scattering amplitude, which is necessary to estimate the cutoff, requires information on curvature tensors. A more comprehensive analysis can be found in Refs.~\cite{He:2023geometry, He:2023vlj}, where the authors discuss the cutoff scale in a frame-independent way.  

Expanding around \( \phi_i = 0 \), which is the case for Higgs-Modular inflation both during and after inflation, the Ricci scalar \( \mathbb{R} \) can be computed as  
\begin{align}
\mathbb{R}=\frac{8\xi}{\Mpl^2}\left(9\xi+\frac{5}{(2\tau_2)^{k_H}}\right)-\frac{4a^2}{9\Mpl^2}\left(18+36k_H+5k_H^2(-3+\pi\tau_2)^2\right),
\end{align}
where we have reinstated the Planck scale. The cutoff scale \( \Lambda \) could then be estimated as \( \Lambda \sim |\mathbb{R}|^{-1/2} \)~\cite{Mikura:2021clt}. We observe that the non-minimal coupling \( \xi \) lowers the cutoff scale, as in the Higgs inflation model, while the modulus \( \tau_2 \) also contributes, as expected, since \( \tau_2 \) is related to the species scale. These contributions cannot be eliminated without fine-tuning, and therefore, the unitarity issue of Higgs inflation appears to persist.  

To examine this in more detail, let us explicitly check the values of the cutoff scale during and after inflation (at the vacuum) and compare them to the typical energy scales in each case.  

During inflation, the cutoff scale should be compared to the Hubble scale \( H \), since particles are excited by gravitational interactions, with their typical energy characterized by the Hubble scale. Using the parameter sets employed in this paper for Case I (Modular-like) and Case II (Higgs-like), we find that the ratio of the Hubble scale to the cutoff scale \( \Lambda_{\rm{inf}} \) is given by  
\begin{align}
\frac{H}{\Lambda_{\rm{inf}}} \sim  
\begin{cases}
5\times 10^{-3} \quad &{\rm{(Case\ I)}} \\
0.1 \quad &{\rm{(Case\ II)}}
\end{cases}
\end{align}
indicating that there is no unitarity violation during inflation in either case.  

Next, we consider the post-inflationary phase during reheating. In this case, the cutoff scale \( \Lambda_{\rm{vac}} \) should be compared to \( V_{\rm{end}} \), the potential energy at the end of inflation, since excited particles can acquire momenta of order \( k \sim V_{\rm{end}}^{1/4} \) during preheating~\cite{Ema:2016dny}. Performing an order-of-magnitude estimation, we find  
\begin{align}
\frac{V_{\rm{end}}^{1/4}}{\Lambda_{\rm{vac}}} \sim  
\begin{cases}
0.8 \quad &{\rm{(Case\ I)}} \\
94 \quad &{\rm{(Case\ II)}}
\end{cases}
\end{align}
suggesting that Case II may suffer from unitarity violation at the vacuum. However, whether efficient preheating occurs in our setup requires further investigation, as metric Higgs inflation and Palatini Higgs inflation exhibit entirely different preheating structures. We leave this issue for future work.

\section{Perturbative Reheating}\label{Sec4}
In this section, we discuss post-inflationary dynamics, specifically reheating. Reheating is important not only for connecting inflation to standard Big Bang cosmology but also for particle production, including dark matter, as well as gravitational wave generation.

\begin{figure}[t]
 \begin{minipage}{0.5\hsize}
  \begin{center}
   \includegraphics[width=60mm]{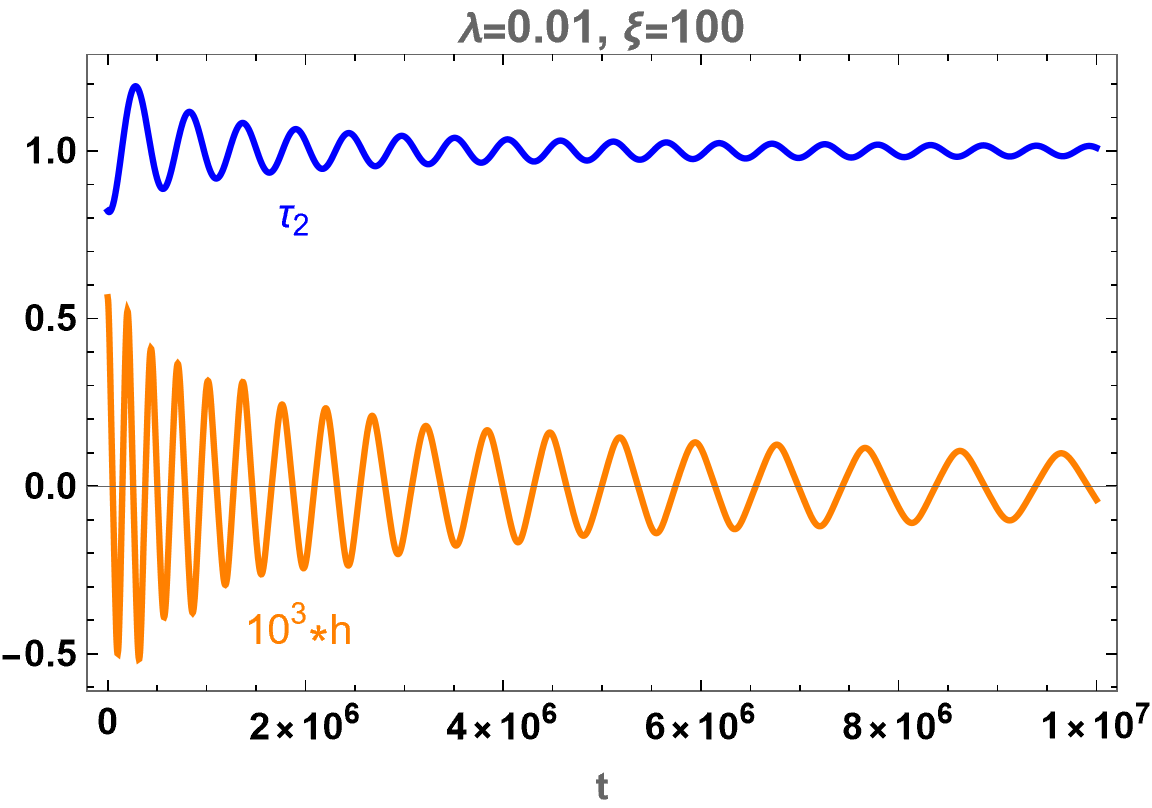}
  \end{center}
 \end{minipage}
 \begin{minipage}{0.5\hsize}
  \begin{center}
   \includegraphics[width=60mm]{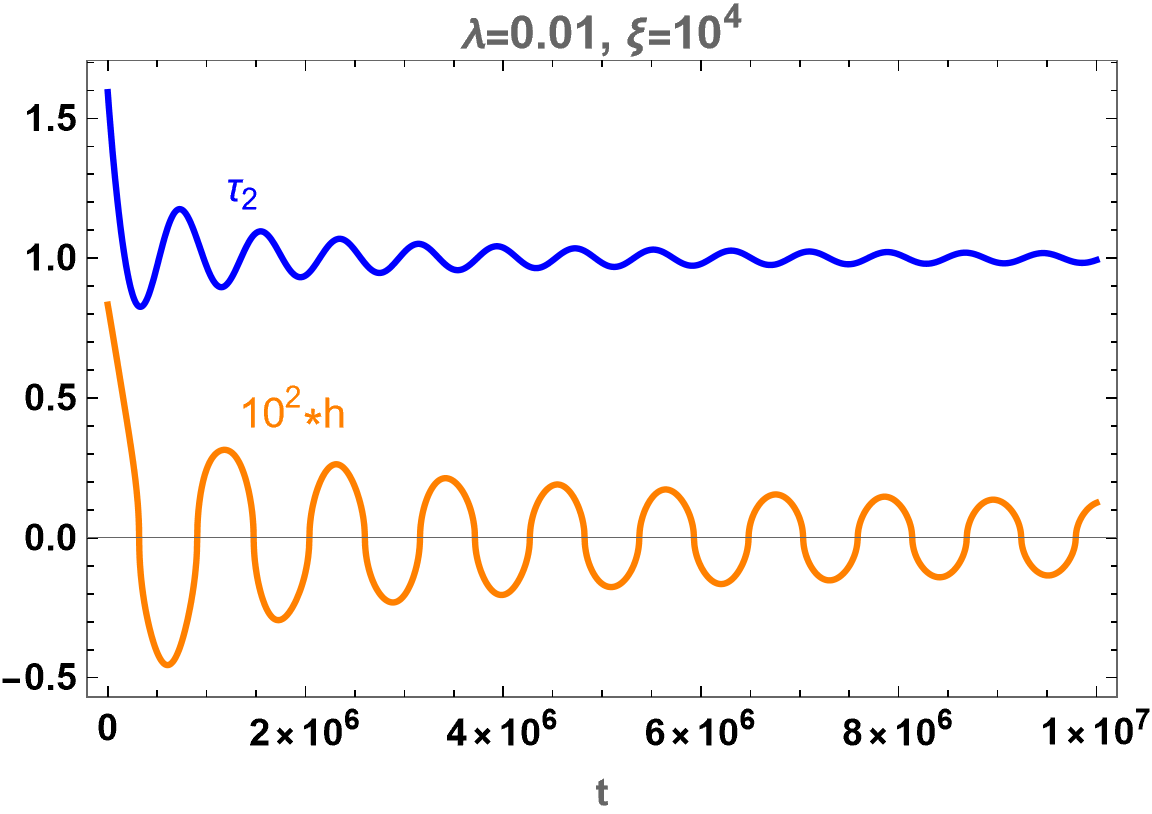}
  \end{center}
 \end{minipage}

\caption{The time evolution of modulus \( \tau_2 \) (blue) and Higgs \( h \) (orange) after inflation end.}
  \label{fig:Osc}
\end{figure}

In the following discussion, we set \( a = c = 1 \) for simplicity. 
After inflation, \( \tau_2 \) and \( h \) start to oscillate almost independently, following\footnote{This situation differs from the Higgs-\( R^2 \) inflation model~\cite{He:2020qcb,Aoki:2022dzd}, where oscillations of the Higgs and scalaron are correlated due to the cubic interaction~$\phi h^2$ giving a tachyonic mass to $h$ when $\phi<0$.}  
\begin{align}
V\simeq 4V_0 \phi_c^2 +\frac{\lambda}{4}h_c^4,   \label{V_osci}
\end{align}
where \( \phi_c \) and \( h_c \) are the canonically normalized fields around the vacuum \( (\tau_2,h)\sim (1,0) \),
\begin{align}
 \tau_2=1+2 \phi_c, \quad h=2^{k_H/2} h_c,   
\end{align}
and we simply denote them as \( \phi \) and \( h \) in the following, omitting the subscript.
Note that \( \phi \) (\( h \)) oscillates with a quadratic (quartic) potential.  
We numerically solved the equations of motion~\eqref{EOM1}-\eqref{EOM3} from the end of inflation \( (\epsilon=1) \) onward. See Fig.~\ref{fig:Osc}. In both cases with \( \xi=10^2 \) and \( \xi=10^4 \) with $\lambda=0.01$ fixed, one can see that the inflatons \( (\tau_2,h) \) start oscillating around \( (\tau_2,h)\sim (1,0) \) after inflation.

While the inflatons are oscillating, they decay into the thermal plasma, reheating the universe. Before going to the reheating analysis of the Higgs-Modular system, let us first review the results of pure modular inflation.

\subsection{Pure modular inflation}
The reheating process in pure modular inflation, where the Higgs field plays no role, has been discussed in Ref.~\cite{Ding:2024euc}. There, it is shown that the modulus \( \phi \) predominantly decays into right-handed neutrinos \( N \) (with mass \( M \))—if kinematically allowed—in most of the parameter space via  
\begin{align}
\mathcal{L}_{\phi NN}= \frac{\lambda_N M}{2\Mpl}  \phi N N,\label{L_pNN}
\end{align}
where \( \lambda_N \) is taken to be a real constant, and we keep only the first generation of neutrinos. A detailed and more general analysis including other possible  decay channels such as three-body decay with Yukawa couplings can be found in Ref.~\cite{Ding:2024euc}.  

The decay rate of the modulus \( \phi \) into right-handed neutrinos \( N \), as derived from Eq.~\eqref{L_pNN}, is given by  
\begin{align}
\Gamma\left(\phi \rightarrow N N\right) = \frac{\lambda_N^2 m_\phi}{16 \pi} \left(\frac{M}{\Mpl}\right)^2 \left(1 - \frac{4M^2}{m^2_\phi}\right)^{3/2}, \label{G_phiNN}
\end{align}
where \( m^2_{\phi} = 8V_0\Mpl^2 \) is the modulus mass at the vacuum. Explicitly inserting numerical values, the order of magnitude can be estimated as   
\begin{align}
\Gamma\left(\phi \rightarrow N N\right)  \sim 10^{-6} \, \text{GeV} \times \left(\frac{M}{10^{10} \, \text{GeV}}\right)^2 \times \left(\frac{V_0}{10^{-12}}\right)^{1 / 2},
\end{align}
for \( \lambda_N \sim \mathcal{O}(1) \). From this, the reheating temperature in the pure modular inflation scenario is given by~\cite{Ding:2024euc}  
\begin{align}
T_{\rm{reh, Pure\ Modular}} \sim \sqrt{\Gamma_\phi \Mpl} \sim 10^{6} \, \text{GeV} \times \left(\frac{M}{10^{10} \, \text{GeV}}\right) \times \left(\frac{V_0}{10^{-12}}\right)^{1/4}, \label{P_T_reh}    
\end{align}
which leads to a low reheating temperature.  

\subsection{Higgs-Modular inflation}
Let us move to Higgs-Modular system. In our case, in addition to the above decay processes~\eqref{G_phiNN}, there are two additional contributions.  

First, the modulus can also decay into Higgs quanta \( \delta h \) through derivative couplings~\eqref{cov_h} due to the nonzero Higgs charge under the modular symmetry. The decay rate can be estimated as  
\begin{align}
\Gamma(\phi \rightarrow \delta h \delta h) = \frac{g^2 m_\phi^3}{16 \pi \Mpl^2} \left(1-\frac{4 m_h^2}{m_\phi^2}\right)^{1 / 2},
\end{align}
where \( m_h^2 \equiv \lambda h^2/2 \) and \( g = k_H\pi/3 \). This decay channel opens at a late stage of reheating, after the Higgs oscillation amplitude becomes small. Although the rate is Planck-suppressed—again reflecting the nature of the modulus field—it is larger than the decay rate given in Eq.~\eqref{G_phiNN}:  
\begin{align}
\Gamma(\phi \rightarrow \delta h \delta h) \sim 1 \, \text{GeV} \times \left(\frac{V_0}{10^{-12}}\right)^{3 / 2},\label{G_phh}
\end{align}
which leads to an increase in the reheating temperature.

Second, the Higgs field is involved as a part of the inflaton system and can also decay into Standard Model (SM) particles. For example, let us consider Higgs decay into fermion pairs, \( h\rightarrow f+\bar{f} \), with a Yukawa coupling \( \sim y h\bar{f}f/\sqrt{2} \). The situation is similar for the decay to gauge bosons. Taking into account that the Higgs potential is quartic around the minimum~\eqref{V_osci}, the corresponding decay rate can be calculated as~\cite{Garcia:2020wiy}  
\begin{align}
\Gamma(h \rightarrow f\bar{f})=\frac{y^2}{4 \pi \rho_h^{1/2}} \sum_{k=0}^{\infty} (k \omega)^3\left|\mathcal{P}_k\right|^2\left[1-\left(\frac{2 m_f}{k \omega}\right)^2\right]^{3/2},   \label{G_h} 
\end{align}
where we decompose the Higgs condensate \( h(t) \) into a slowly varying part \( \mathsf{h}(t) \), which includes the redshift, and a rapidly oscillating part \( \mathcal{P}(t) \),  
\begin{align}
h(t)=\mathsf{h}(t) \cdot \mathcal{P}(t).    
\end{align}
The latter can be further decomposed into a Fourier series as  
\begin{align}
\mathcal{P}(t)=\sum_{k=-\infty}^{\infty} \mathcal{P}_k e^{-i k \omega t},    
\end{align}
where \( \omega \) is the oscillation frequency related to the (time-dependent) Higgs mass by  
\begin{align}
\omega=m_h \sqrt{\frac{2\pi }{3}} \frac{\Gamma\left(\frac{3}{4}\right)}{\Gamma\left(\frac{1}{4}\right)}\sim 0.5 \times m_h,  \quad {\rm{where}} \quad m^2_h=\frac{\lambda}{2}\mathsf{h}^2. 
\end{align}
Finally, we define the (approximate) energy densities of the modulus (\( \phi \)) and the Higgs (\( h \)) by averaging over one oscillation period,  
\begin{align}
\rho_\phi\equiv  \left\langle\frac{1}{2}\dot{\phi}^2+\frac{1}{2}m^2_\phi\phi^2\right\rangle, \quad 
\rho_h\equiv \left\langle \frac{1}{2}\dot{h}^2+\frac{1}{2}m^2_hh^2\right\rangle. \label{def_rho}
\end{align}
For our benchmark value of the Higgs quartic coupling, \( \lambda=0.01 \), which we frequently use in this paper, we numerically find the energy densities at the end of inflation as  
\begin{align}
& \rho_{\phi, \text { end }}/\Mpl^4\sim 10^{-12},\quad \rho_{h, \text { end }}/\Mpl^4\sim 10^{-16} \quad {\rm{for}}\quad \xi=10^2,\label{xi=100}\\
& \rho_{\phi, \text { end }}/\Mpl^4\sim  \rho_{h, \text { end }}/\Mpl^4\sim 10^{-11} \quad {\rm{for}}\quad \xi=10^4.\label{xi=10000}
\end{align}
Therefore, the fraction of the Higgs energy density is small for small non-minimal coupling \( \xi \) (Modular-like) but becomes comparable for large \( \xi \) (Higgs-like).

Now, looking at the kinematic factor in Eq.~\eqref{G_h}, the decay channel opens for Yukawa couplings satisfying \( y < 4k\sqrt{\lambda} \), regardless of the Higgs oscillation amplitude. This condition is specific to inflaton oscillations with a quartic potential~\cite{Garcia:2020wiy,Ahmed:2022tfm,Clery:2023ptm}.  
Thus, only fermions with sufficiently small Yukawa couplings, \( y < 4k\sqrt{\lambda} \), can be produced via this process. However, at the same time, the overall decay rate is suppressed by \( y^2 \).  
For instance, neglecting higher-order anharmonic effects, the Higgs decay rate can be roughly estimated as
\begin{align}
\Gamma(h \rightarrow f f) \sim 10^7 \,\text{GeV} \times \left(\frac{y}{0.1}\right)^2 \times \left(\frac{\lambda}{0.01}\right) \times \left(\frac{\mathsf{h}}{10^{-4} M_{\mathrm{Pl}}}\right). \label{G_hff}
\end{align}

Therefore, in the Higgs-Modular inflation scenario, new decay channels open (Eqs.~\eqref{G_phh} and \eqref{G_hff}), significantly modifying the reheating process.  
As discussed in Ref.~\cite{Aoki:2022dzd}, in a two-field reheating scenario, where the total energy density of the inflatons can be approximated as a sum-separable form, \( \rho_{\rm{total}}\simeq \rho_\phi+\rho_h \),  
the reheating temperature can be estimated as
\begin{align}
T_{\mathrm{reh,  Higgs-Modular}}^4=\frac{72 M_{\mathrm{Pl}}^2}{5 \pi^2 g_{\mathrm{reh}}}\left(\frac{\Gamma_\phi \rho_{\phi, \text { end }}+\Gamma_h \rho_{h, \text { end }}}{\rho_{{\rm{total}}, \text { end }}}\right)^2,    
\end{align}
where \( g_{\mathrm{reh}} \) is the effective number of relativistic degrees of freedom at the end of reheating, and the decay rates \( \Gamma_\phi \) and \( \Gamma_h \) are given in Eqs.~\eqref{G_phh} and \eqref{G_hff}.  
This leads to the approximate reheating temperatures:
\begin{align}
T_{\rm{reh, Higgs-Modular}} \sim \left\{\begin{array}{l}
\sqrt{\Gamma_\phi M_{\mathrm{Pl}}} \sim  10^{10} \,\text{GeV},\quad \xi=10^2 \quad({\rm{Modular-like}}), \\
\sqrt{\Gamma_h M_{\mathrm{Pl}}} \sim 10^{12} \,\text{GeV},\quad \xi=10^4 \quad({\rm{Higgs-like}}).
\end{array}\right.
\end{align}
where we have used the energy budget estimates from Eqs.~\eqref{xi=100} and \eqref{xi=10000}.  

In the Modular-like case (\( \xi=10^2 \)), although the modulus decay rate \( \Gamma_\phi \) is much smaller than the Higgs decay rate \( \Gamma_h \), the Higgs contribution is suppressed by the energy density ratio \( \rho_{h, \text { end }}/\rho_{\phi, \text { end }} \), so that the reheating temperature is primarily determined by modulus decay.  
Notably, despite being in a Modular-like regime, the reheating temperature is still higher than in the pure modular inflation case~\eqref{P_T_reh} due to the additional decay channel~\eqref{G_phh} induced by the nonzero modular weight of the Higgs.  
This difference may lead to distinct phenomenological consequences.  

On the other hand, in the Higgs-like case (\( \xi=10^4 \)), there is no suppression of the energy density ratio since \( \rho_{h, \text { end }}\sim \rho_{\phi, \text { end }} \) in this case, and the reheating temperature is almost entirely determined by Higgs decay rate leading to high reheating temperature.

\section{Conclusions and Discussions}
\label{sec:con}

In this paper, we discussed the inflationary dynamics and its observational predictions in a system where both the modulus and the Standard Model Higgs boson contribute to inflation, which we call the Higgs-Modular inflation model. This scenario naturally arises when the modulus plays the role of the inflaton (the so-called modular inflation scenario) and the Higgs field is charged under the underlying modular (or \( SL(2,\mathbb{Z}) \)) symmetry. Indeed, there is evidence that the Higgs field is charged under modular symmetry, as is known in string compactifications (e.g.,~\cite{LopesCardoso:1994is,Antoniadis:1994hg,Brignole:1995fb}) and also in the context of modular flavor models (e.g.,~\cite{Ishiguro:2021drk,Ishiguro:2024xph}). 
In such cases, and in the presence of a non-minimal coupling of the Higgs field to gravity ($\xi \neq 0$), which is naturally induced by renormalization group (RG) effects, the Higgs field inevitably participates in the inflationary dynamics, and its effects should be properly taken into account. In particular, due to Higgs-modulus interactions governed by \( SL(2,\mathbb{Z}) \) symmetry, we found that the inflationary dynamics become nontrivial in certain regions of parameter space, leading to two-field inflation. Our system interpolates between pure modulus inflation and pure Higgs inflation in some extreme limits.  

We also derived inflationary observables and compared them with observational data, including the recent ACT results. The inflationary dynamics, and hence the observational predictions, strongly depend on the ratio of the Higgs quartic coupling to the Higgs non-minimal coupling to gravity when the Higgs charge under the modular symmetry is nonzero. We classified the scenarios into modular-like and Higgs-like cases (Case I and Case II, respectively, in the main text), which correspond to relatively small and large non-minimal couplings. While the observational prediction for Case I is almost degenerate with those of pure Higgs and pure modulus inflation scenarios, we found that Case II predicts unique values for observables due to nontrivial Higgs-modulus interactions. In particular, the predicted spectral index is slightly higher than that of pure Higgs inflation, which is favored by ACT data.  

Additionally, we discussed the unitarity issue, which is present in the pure Higgs inflation scenario, in our setup. Although further analysis of preheating in this model is necessary, we observed that Higgs-Modular inflation may generally require UV completion, which will be left for future work. 

Furthermore, we examined reheating after inflation and compared it with pure modular inflation. We found that in the Higgs-Modular inflation scenario, additional decay channels become available due to Higgs-modulus couplings and Higgs-Standard Model interactions. We also found that, due to these additional interactions, Higgs-Modular inflation generally predicts a higher reheating temperature than pure modular inflation across most of the parameter space. This increase in reheating temperature may have significant implications for post-inflationary phenomenology, such as dark matter production during and after reheating.  

There are several future directions to explore.  

First, a more detailed analysis of primordial perturbations would be interesting. As we found, our two-field system (Higgs and modulus) naturally leads to a curved trajectory, which could potentially source primordial non-Gaussianity in the curvature mode. In particular, if the direction orthogonal to the inflaton trajectory (the isocurvature mode) is not too heavy (around the Hubble scale), we may expect direct information about such heavy particles to be imprinted on non-Gaussianity, a phenomenon known as the cosmological collider signal~\cite{Chen:2009zp, Baumann:2011nk, Noumi:2012vr, Arkani-Hamed:2015bza}. The various conditions that need to be satisfied for the cosmological collider mechanism to be effective have been clarified in Ref.~\cite{Aoki:2024jha}, and it would be interesting to check them in our setup. Additionally, as briefly mentioned in the main text, the dynamics and perturbations of the axion \(\tau_1\), which we have ignored in this work, may play an important role in certain cases. Investigating these multifield effects on primordial perturbations could be key to distinguishing between different extended versions of Higgs inflation.  

Second, in this work, we assumed that there are no significant preheating effects and simply considered a scenario in which reheating is completed via perturbative decays of inflatons. However, this assumption requires further verification. In the case of the Higgs-\(R^2\) model, it has been recently pointed out that preheating effects can become significant, especially for large non-minimal coupling \( \xi \) (a Higgs-like situation in our terminology)~\cite{Kim:2025ikw}. In some regions of parameter space in our model, preheating effects may need to be taken into account, making it crucial to understand the correct post-inflationary dynamics. Furthermore, this is important for understanding unitarity violation after inflation, as mentioned above.  

We leave these questions for future work.


\acknowledgments
We would like to thank Minxi He for useful discussion. S.A. is supported by the Japan Science and Technology Agency (JST) as part of Adopting Sustainable Partnerships for Innovative Research Ecosystem (ASPIRE),  Grant No. JPMJAP2318. This work was supported in part by JSPS KAKENHI Grant Numbers JP25H01539 (H.O).

\bibliography{references}{}
\bibliographystyle{JHEP}

\end{document}